\documentstyle[aps,epsfig,preprint]{revtex}

\begin{document}
\draft
\title{Controlling the exchange interaction using the spin-flip transition of antiferromagnetic spins in Ni$_{81}$Fe$_{19}$ / $\alpha$-Fe$_2$O$_3$}
\author{Joonghoe Dho, C. W. Leung, Z. H. Barber, and M. G. Blamire\cite{Blamire}}
\address{Department of Materials Science and Metallurgy, University of Cambridge,
Pembroke Street, Cambridge CB2 3QZ, UK }
\date{\today}
\maketitle

\begin{abstract}

We report studies of exchange bias and coercivity in ferromagnetic
Ni$_{81}$Fe$_{19}$ layers coupled to antiferromagnetic (AF)
(0001), (11$\bar{2}$0), and (11$\bar{0}$2) $\alpha$-Fe$_2$O$_3$
layers. We show that AF spin configurations which permit spin-flop
coupling give rise to a strong uniaxial anisotropy and hence a
large coercivity, and that by annealing in magnetic fields
parallel to specific directions in the AF we can control either
coercivity or exchange bias. In particular, we show for the first
time that a reversible temperature-induced spin reorientation in
the AF can be used to control the exchange interaction.

\end{abstract}

\narrowtext

The interaction between a ferromagnet (FM) and an antiferromagnet
(AF) across an interface gives rise to an exchange bias
($H_{ex}$), i.e., a shift in the hysteresis loop as well as an
enhanced coercivity ($H_C$) compared with a free FM material
\cite{Meiklejohn}. This exchange bias is fundamental to the
operation of spin valve devices such as magnetic read-heads,
non-volatile memories, and various sensors
\cite{Dieny,Prinz,Prieto}. Despite considerable work by many
groups over the past two decades, the origin of the exchange bias
and the enhanced coercivity are still unclear
\cite{Nogues1,Berkowitz,Borchers,Steadman}. One of the primary
aims of the various models is the reconciliation of the
experimentally observed values of $H_{ex}$ and $H_C$ with
theoretical predictions. Mauri $et$ $al$.\cite{Mauri} predict
experimentally realistic values for $H_{ex}$ on the assumption
that a domain wall parallel to the surface is formed in the AF
layer. The random field model of Malozemoff \cite{Malozemoff}, by
considering interface roughness, qualitatively explained exchange
bias in compensated, but disordered, AF surfaces. Koon \cite{Koon}
demonstrated that it is possible for the FM layer to minimize its
energy when it aligns perpendicularly to the AF easy axis; this
type of perpendicular exchange coupling has become known as
spin-flop coupling because of its similarity to the spin flop
state of an AF material in a magnetic field. However, Schulthess
and Butler \cite{Schulthess} revealed that the spin-flop coupling
alone cannot induce a unidirectional anisotropy, but instead gives
rise to a uniaxial anisotropy which causes an enhanced coercivity.
Experimentally, such spin-flop coupling has been observed in
epitaxial FM/AF systems, such as
Ni$_{80}$Fe$_{20}$/Fe$_{50}$Mn$_{50}$, Co/NiO, Fe$_3$O$_4$/CoO,
and Fe/FeF$_2$
\cite{Jungblut,Dubourg,Ijiri,Nogues2,Moran,Fitzsimmons}.
Nevertheless, a satisfactory understanding is not yet available
because of complications at interfaces which include roughness,
spin structure, and defects. Of particular relevance to the work
presented here, Fitzsimmons $et$ $al$. \cite{Fitzsimmons} have
shown that exchange bias is dependent on the in-plane crystalline
quality, and hence the net spin configuration at the interface, of
an AF layer. The aim of the experiments reported here was to
investigate the exchange bias and coercivity in a system in which
the interfacial AF spin configuration could be controlled and
changed without modifying the structural properties of the AF/FM
interface.

Hematite $\alpha$-Fe$_2$O$_3$ is potentially attractive for
exchange biased applications because of its high Neel temperature
($\sim$680 $^0$C) \cite{Sano,Bae}. Bulk $\alpha$-Fe$_2$O$_3$
undergoes an unusual temperature-controlled transition between two
AF spin configurations - the so-called spin-flip (Morin)
transition at $T_M\sim$260 K. Thus the spin configuration of a
single-crystal $\alpha$-Fe$_2$O$_3$ surface not only depends on
the direction of that surface, but can be changed by altering the
temperature. It has already been identified as an ideal system in
which to study exchange bias in general and spin flop in
particular \cite{Leighton}.

It has been shown that epitaxial $\alpha$-Fe$_2$O$_3$ films on
$\alpha$-Al$_2$O$_3$ substrates have spin-flip transition
temperatures which depend on the crystal orientation
\cite{Fujii,Gota}: the $T_M$ of (11$\bar{2}$0)
$\alpha$-Fe$_2$O$_3$ is similar to that of the bulk material,
while (0001) $\alpha$-Fe$_2$O$_3$ films do not show a spin-flip
transition above 2.5 K. In contrast, the spin-flip transition of
(1$\bar{1}$02) $\alpha$-Fe$_2$O$_3$ films is increased to about
400 K; in this case the AF spins lying within the film plane above
$T_M$ flip to the out-of-plane direction below $T_M$. These
changes are associated with lattice strain caused by the lattice
mismatch between $\alpha$-Fe$_2$O$_3$ and $\alpha$-Al$_2$O$_3$
($\sim$5.5 \%).

In this Letter, we report exchange bias and coercivity in FM
layers coupled with epitaxial $\alpha$-Fe$_2$O$_3$ layers. In
contrast to previous experiments which have compared the exchange
interaction associated with different $fixed$ spin orientations
associated with different AF crystal faces \cite{Nogues2}, we show
for the first time that a change of AF spin orientation across a
$single$ interface is directly reflected in a modified exchange
interaction.

Epitaxial $\alpha$-Fe$_2$O$_3$ films were grown on (0001),
(11$\bar{2}$0), and  (1$\bar{1}$02) $\alpha$-Al$_2$O$_3$
substrates by pulsed laser deposition (PLD) with a substrate
temperature of 700 $^0$C and oxygen pressure of 20 mTorr. In order
to fabricate the films under identical conditions, three
substrates were loaded side by side for simultaneous deposition.
The 50 nm thick $\alpha$-Fe$_2$O$_3$ films were transferred into
an ultra high vacuum dc sputtering chamber and a 5 nm
Ni$_{81}$Fe$_{19}$(NiFe) film was deposited in a magnetic field of
250 Oe at 295 K.

X-ray diffraction (XRD) measurements showed that all three
$\alpha$-Fe$_2$O$_3$ growth directions yielded films with
excellent crystallinity: a full width half maximum of $<$ 0.08$^0$
and $\sim$0.9$^0$ in the rocking curve and in the $\phi$ scan,
respectively. The $rms$ surface roughness measured by atomic force
microscopy was about 0.5 nm in each case. Therefore, effects
caused by extrinsic factors such as roughness and defects should
be virtually same for all three samples, and thus differences in
the $H_{ex}$ and the $H_C$ should depend only on the spin
structure of $\alpha$-Fe$_2$O$_3$ at the surface. The strains
observed in our films are similar to those reported by Fujii et
al. \cite{Fujii} and so we expect a similar change in the Morin
temperature.

The NiFe/$\alpha$-Fe$_2$O$_3$ samples were measured in a variable
temperature vibrating sample magnetometer (VSM). Fig. 1 shows the
temperature-dependent magnetic hysteresis loops of as-prepared
samples for the different crystal orientations; each panel shows
magnetization data collected for two orthogonal in-plane
directions (see Fig. 3). For the (0001) orientation, the
hysteresis loop is essentially independent of the temperature and
the in-plane field direction: it shows minimal $H_{ex}$, and a
coercive field of about 22 Oe. In contrast, the (11$\bar{2}$0)
$\alpha$-Fe$_2$O$_3$ system shows large changes between 25 and 295
K. Fig. 2 shows how the magnetization and saturation field of
(11$\bar{2}$0) $\alpha$-Fe$_2$O$_3$ / NiFe for two in-plane
directions depends on temperature; it is clear that the easy axis
rotates by 90$^0$ over the temperature range of the experiment.
Finally, hysteresis loops of the NiFe on (1$\bar{1}$02)
$\alpha$-Fe$_2$O$_3$ / NiFe for two in-plane directions at 295 K
are virtually identical except a slight shift, but at 380 K there
is a clear difference for the two in-plane directions, i.e. easy
and hard magnetization axes.

Fig. 3 shows the crystal-direction dependence of the surface spin
configuration of $\alpha$-Fe$_2$O$_3$\cite{Fujii,Gota,Fuller}; the
NiFe spin will always lie within the film plane because of large
shape anisotropy. For spin-flop coupling to give rise to a strong
uniaxial anisotropy, the FM spins must align perpendicular to the
AF spins \cite{Schulthess}. Accordingly, if the spin-flip
transition of $\alpha$-Fe$_2$O$_3$ at $T_M$ results in a change of
the in-plane spin direction, this should be reflected by a change
in the easy axis of the NiFe. This expectation is consistent with
our results for the (11$\bar{2}$0) and (1$\bar{1}$02)
$\alpha$-Fe$_2$O$_3$, as seen in Figs. 1 and 2: the 90$^0$
rotation of the easy axis of NiFe on the (11$\bar{2}$0)
$\alpha$-Fe$_2$O$_3$ is associated with the spin-flip transition
of $\alpha$-Fe$_2$O$_3$ from the $ab$ plane to the $c$-axis
(albeit associated with a reduced $T_m$ which is typical for 40-50
nm length scales in $\alpha$-Fe$_2$O$_3$ \cite{Fujii,Vasquez}),
while the FM spins on (1$\bar{1}$02) $\alpha$-Fe$_2$O$_3$ at room
temperature have no preferential orientation because all in-plane
directions equally satisfy a spin-flop coupling condition. Upon
warming, however, a preferential direction appears within the
plane as the AF spins flip to one of the in-plane
directions.\cite{Fujii} The substantially lower coercivity for the
NiFe on (0001) $\alpha$-Fe$_2$O$_3$ is also consistent with this
picture since it has an uncompensated surface at all temperatures,
and so cannot generate spin-flop coupling. It is important to note
that this uncompensated surface gave negligible exchange bias
following any annealing procedure in contradiction to simple
models for such systems; this may be a consequence of the small,
but finite roughness in any practical sample. Although this may
appear surprising, it is consistent with previous results in the
exchange biased epitaxial system Fe/FeF$_2$, which showed zero
exchange bias for an uncompensated AF surface \cite{Nogues2}.

If the intrinsic anisotropy energy of a FM layer is negligible,
the total energy per unit area in an exchange coupled FM/AF system
can be expressed as \cite{Hu}

\begin{equation}
E=-J_1 cos \theta - J_2 sin^2 \theta + K_{AFM}sin^2 \phi
\end{equation}

where $J_1$ and $J_2$ are respectively a direct (parallel)
coupling constant and a spin-flop (perpendicular) coupling
constant; $\theta$ and $\phi$ are the angles between the FM spin
and the AF spin directions, and the AF spin and the AF anisotropy
axis; $K_{AFM}$ is the anisotropy constant of the AF layer. The
lowest energy state is thought to be a spin flop like state
($\theta$ = 90$^0$, $\phi$ = 0$^0$ (see $T$ $>$ $T_m$ in Fig.
3(b),(c)). The form of the spin flop coupling is comparable with
the classical uniaxial anisotropy energy, and thus the coercivity
is mainly dependent on the second term of (1). If we associate
exchange bias with a domain wall formed in the AF layer
\cite{Borchers,Malozemoff,Nowak}, its stability is determined by a
competition between a decrement of direct coupling energy and an
increment of the AF anisotropy energy. From equation (1), we
expect that a magnetic field annealing (MFA) process perpendicular
to the AF spin direction will stabilize the spin-flop coupling,
and in turn it will enhance the coercivity. On the other hand, a
MFA process parallel to the AF spin direction should induce an
exchange bias because it enhances the direct coupling and
suppresses the spin flop coupling.

We performed a series of experiments in which MFA was performed
under 10 kOe for 15 minutes at 200 $^0$C, and the samples were
cooled down to room temperature in the magnetic field. The
(11$\bar{2}$0) and (1$\bar{1}$02) $\alpha$-Fe$_2$O$_3$ films with
a compensated surfaces showed distinctive MFA effects; Fig. 4
shows the hysteresis loops of NiFe on (1120) $\alpha$-Fe$_2$O$_3$
before and after MFA. When the MFA was performed perpendicular to
the AF spin direction (Fig. 4(d)), the exchange bias showed no
change, but the coercivity approximately doubled (Fig. 4 (b)). In
contrast, $H_{ex}$ of order 80 Oe was induced along the hard axis
when the MFA was performed parallel to the AF spin direction (Fig.
4 (c)). A similar exchange bias along the hard axis has been
observed in the epitaxial Fe/FeFe$_2$ system.[17,18] Thus MFA with
a configuration of Fig. 4(e) enhances the direct coupling in the
equation (1), and in turn it induces the $H_{ex}$ along the hard
axis.

Finally, we applied MFA to the NiFe on (1$\bar{1}$02)
$\alpha$-Fe$_2$O$_3$. A large exchange bias of 80-100 Oe was
induced by the MFA along all directions within the plane, as seen
in Fig. 5. If (1$\bar{1}$02) $\alpha$-Fe$_2$O$_3$ has the ideal
spin structure of Fig. 3(c) below $T_M$, the NiFe should have
shown no exchange bias because the spin flop coupling is dominant.
On the contrary, the large exchange bias suggests that the AF
spin-flip transition to the out-of-plane direction during the
cooling process of MFA is frustrated at the interface because of
in-plane FM spins. In Figure 5 (c), the temperature dependent
exchange bias clearly shows an anomaly at $T_*$, which agrees with
the $T_M$ of (1$\bar{1}$02) $\alpha$-Fe$_2$O$_3$ in previous
report \cite{Fujii}. We conjecture that the MFA assisted by FM
spins enhances the direct coupling leading the formation of a
domain wall in the AF layer, and thus the exchange bias is induced
along all in-plane directions.

The magnitude of exchange bias depends on the time and the
temperature of the MFA; the high Neel temperature of
$\alpha$-Fe$_2$O$_3$ might be expected to limit the effectiveness
of low-temperature MFA. In the present study, the largest value of
$H_{ex}$ at room temperature was about 100 Oe for the NiFe/
(1$\bar{1}$02) $\alpha$-Fe$_2$O$_3$ film. If we simply assume that
the effective interface exchange energy corresponds to a net
direct exchange energy, the energy relation is $J_1=M_FH_{ex}t_F$,
where $M_F$ and $t_F$ are the saturation magnetization and
thickness of the FM layer, respectively. Therefore, the estimated
direct coupling constant $J_1$ is about 0.04 erg/cm$^2$ at room
temperature; this value is similar to that in NiFe/NiO
\cite{Shang}. On the other hand, the maximum difference of the
coercivity between NiFe/ (0001) $\alpha$-Fe$_2$O$_3$ film and
NiFe/ (11$\bar{2}$0) or (1$\bar{1}$02) $\alpha$-Fe$_2$O$_3$ film
was about 180 Oe at room temperature. Because such a difference of
the coercivity is mainly due to the spin-flop coupling, the
spin-flop coupling $J_2$ can be estimated to be roughly 0.07
erg/cm$^2$.

In summary, we have shown for the first time that a change of spin
orientation in an AF material is directly reflected in a modified
exchange interaction. We have also demonstrated that magnetic
field annealing parallel to specific directions in the AF can
alternatively modify either the coercivity or the exchange bias,
in agreement with the Schulthess and Butler\cite{Schulthess} and
Malozemoff models\cite{Malozemoff} respectively, demonstrating
that these pictures of exchange bias are not mutually exclusive,
but are part of a wider picture. This systematic study should
improve the understanding of fundamental origins of exchange bias
and the coercivity in exchange biased systems.

This work was supported by the Engineering and Physical Sciences
Research Council (EPSRC) and by the Post-doctoral Fellowship
Program of Korea Science and Engineering Foundation (KOSEF). We
would like to thank Bryan Hickey and Mary Vickers for valuable
advice.

\newpage

\begin{figure}
\epsfig{file=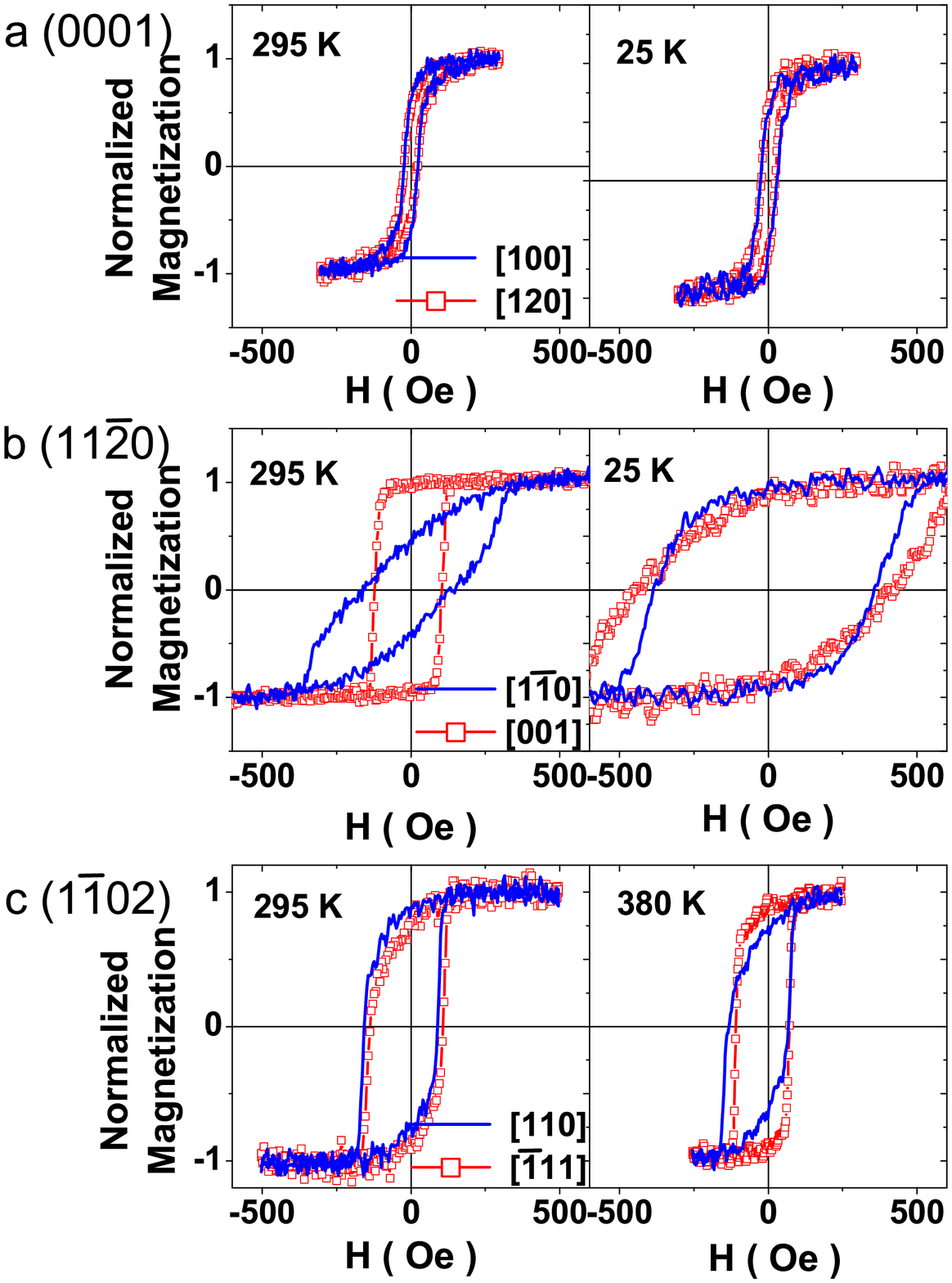,width=14cm} \caption{The magnetic
hysteresis loops of as-prepared Ni$_{81}$Fe$_{19}$ on (a) (0001)
$\alpha$-Fe$_2$O$_3$, (b) (11$\bar{2}$0) $\alpha$-Fe$_2$O$_3$, and
(c) (1$\bar{1}$02) $\alpha$-Fe$_2$O$_3$ at several temperatures. }
\end{figure}

\newpage

\begin{figure}
\epsfig{file=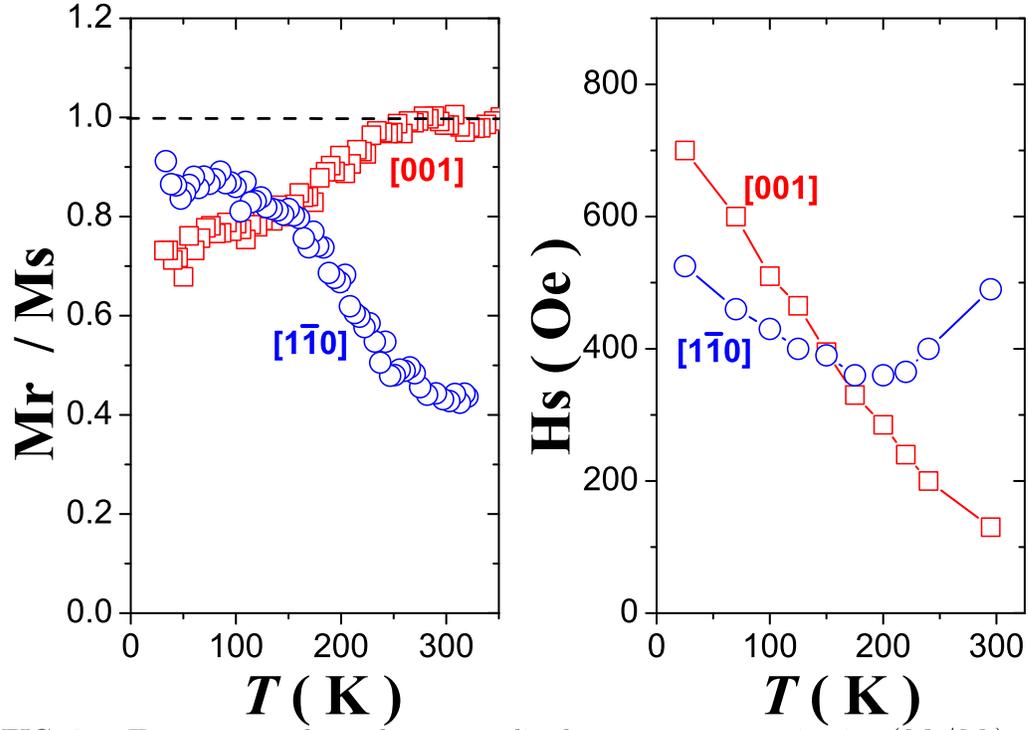, width=14cm} \caption{. Temperature
dependent normalized remanent magnetization ($M_r$/$M_s$) and
saturation field ($H_s$) of as-prepared Ni$_{81}$Fe$_{19}$ on
(11$\bar{2}$0) $\alpha$-Fe$_2$O$_3$ with the in-plane direction. }
\end{figure}

\newpage

\begin{figure}
\epsfig{file=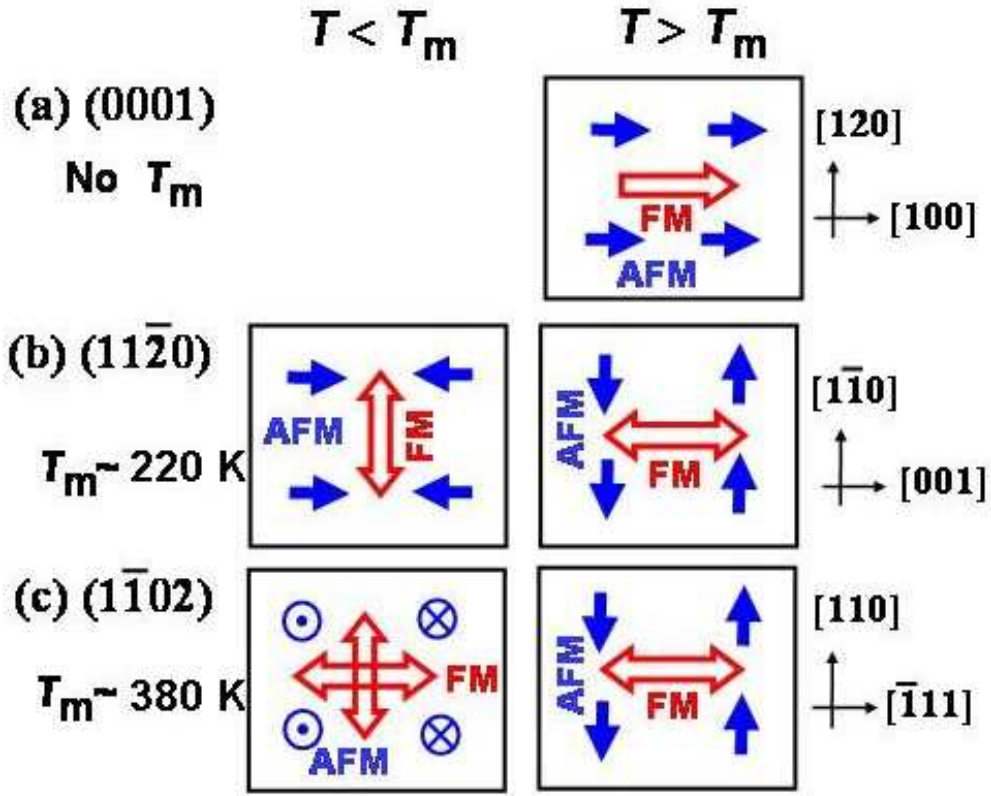,width=14cm} \caption{Schematic surface
spin structures of $\alpha$-Fe$_2$O$_3$ (a) on (0001)
$\alpha$-Al$_2$O$_3$, (b) on (11$\bar{2}$0) $\alpha$-Al$_2$O$_3$,
and (c) on (1$\bar{1}$02) $\alpha$-Al$_2$O$_3$.  }
\end{figure}

\newpage

\begin{figure}
\epsfig{file=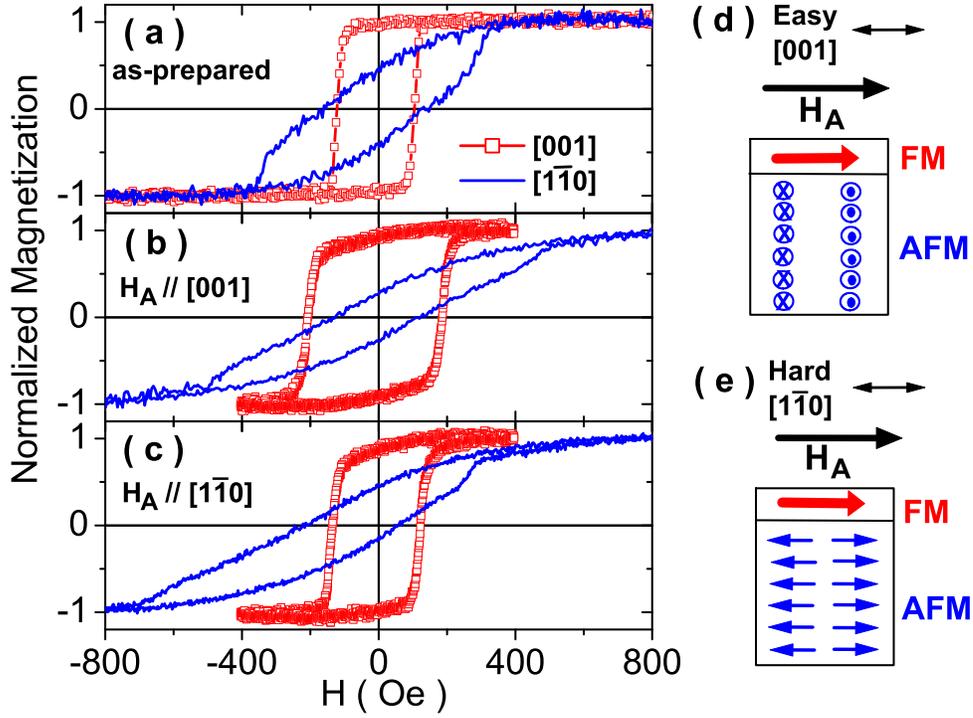,width=14cm} \caption{The room temperature
magnetic hysteresis loops of Ni$_{81}$Fe$_{19}$ on (11$\bar{2}$0)
$\alpha$-Fe$_2$O$_3$ (a) without MFA, (b) with MFA perpendicular
to the AF spin direction, and (c) with MFA parallel to the AF spin
direction. Two schematic MFA configurations were also displayed in
(d) and (e). Here, $H_A$ means the annealing magnetic field.}
\end{figure}

\newpage

\begin{figure}
\epsfig{file=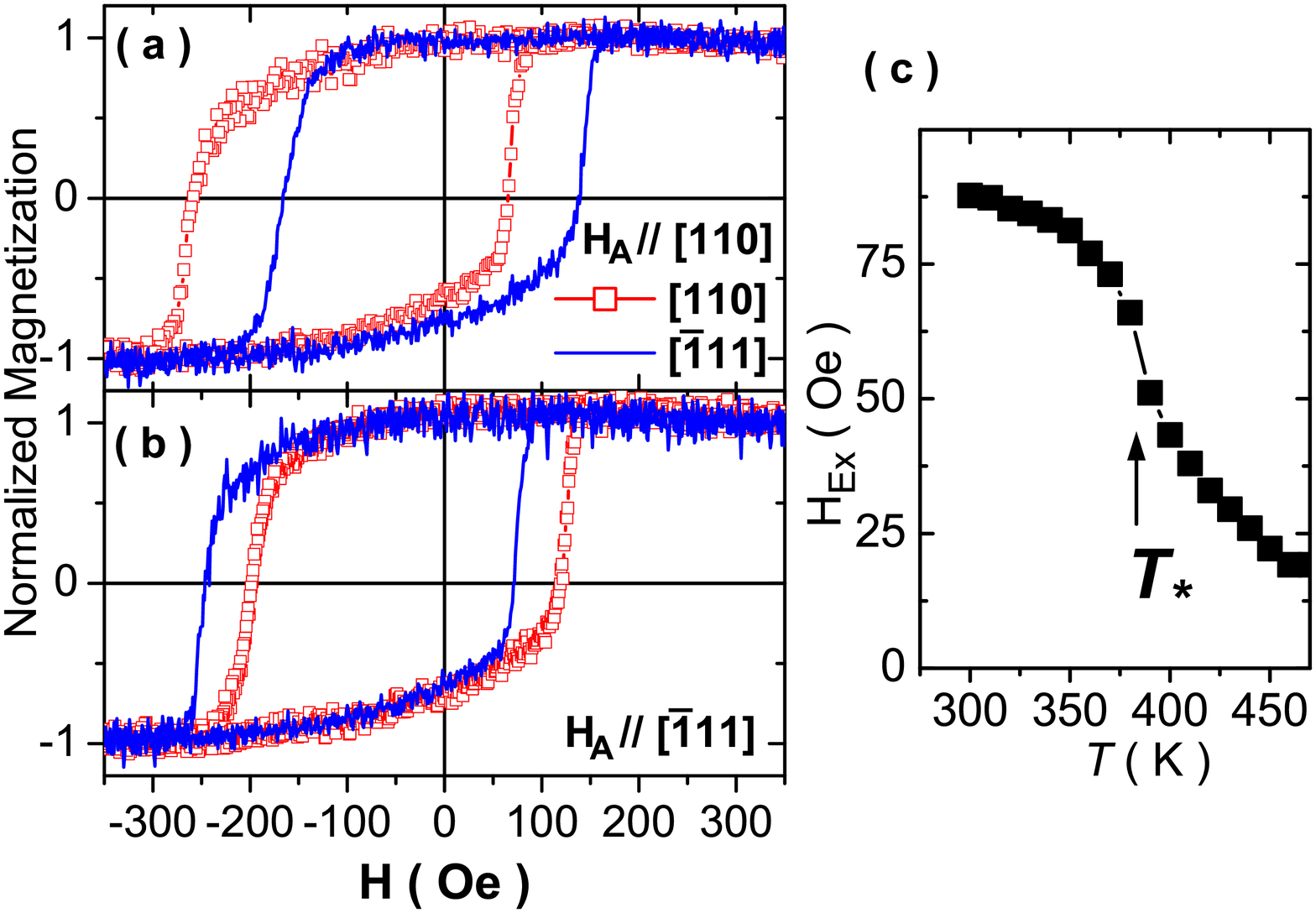,width=14cm} \caption{The room temperature
magnetic hysteresis loops of Ni$_{81}$Fe$_{19}$ on (1$\bar{1}$02)
$\alpha$-Fe$_2$O$_3$ (a) with MFA along the [110] direction, and
(b) with MFA along the [$\bar{1}$11] direction. The temperature
dependent exchange bias H$_{Ex}$ was displayed in (c).  }
\end{figure}



\begin{references}
\bibitem[*]{Blamire} Corresponding author: mb52@cam.ac.uk (e-mail)
\bibitem{Meiklejohn} W. H. Meiklejohn and C. P. Bean, Phys. Rev. {\bf 102}, 1413 (1956).
\bibitem{Dieny} B. Dieny, V.S. Speriosu, S.S.P. Parkin, B.A. Gurney, D.R. Wilhoit and D. Mauri, Phys. Rev. B {\bf 43}, 1297 (1991).
\bibitem{Prinz} G. A. Prinz, Science {\bf 282}, 1660 (1998).
\bibitem{Prieto} J. L. Prieto, N. Rouse, N. K. Todd, D. Morecroft, J. Wolfman, J. E. Evetts and M. G. Blamire, Sensor. Actuat. A : Phys. {\bf 94}, 64 (2001).
\bibitem{Nogues1} J. Nogu\'{e}s and Ivan K. Schuller, J. Magn. Magn. Mater. {\bf 192}, 203(1999)
\bibitem{Berkowitz} A. E. Berkowitz and K. Takano, J. Magn. Magn. Mater. {\bf 200}, 552 (1999).
\bibitem{Borchers}J. A. Borchers, Y. Ijiri, D. M. Lind, P. G. Ivanov, R. W. Erwin, A. Qasba, S. H. Lee, K. V. O'Donovan, and D. C. Dender, Appl. Phys. Lett. {\bf 77}, 4187 (2000).
\bibitem{Steadman} P. Steadman, M. Ali, A. T. Hindmarch, C. H. Marrows, B. J. Hickey, S. Langridge, R. M. Dalgliesh, and S. Foster, Phys. Rev. Lett. {\bf 89}, 077201 (2002).
\bibitem{Mauri} D. Mauri, H. C. Siegmann, P. S. Bagus, and E. Kay, J. Appl. Phys. {\bf 62}, 3047 (1987).
\bibitem{Malozemoff} A. P. Malozemoff, Phys. Rev. B {\bf 35}, 3679 (1987).
\bibitem{Koon} N. C. Koon, Phys. Rev. Lett. {\bf 78}, 4865 (1997).
\bibitem{Schulthess} T. C. Schulthess and W. H. Butler, Phys. Rev. Lett. 81, 4516 (1998).; T. C. Schulthess and W. H. Butler, J. Appl. Phys. {\bf 85}, 5510 (1999).
\bibitem{Jungblut} R. Jungblut, R. Coehoorn, M. T. Johnson, Ch. Sauer, P. J. van der Zaag, A. R. Ball, Th. G. S. M. Rijks, J. ann de Stegge, and A. Reinders, J. Magn. Magn. Mater. {\bf 148}, 300 (1995).
\bibitem{Dubourg} S. Dubourg, N. Negre, B. Warot, E. Snoeck, M. Goiran, J. C. Ousset, and J. F. Bobo, J. App. Phys. {\bf 87}, 4936 (2000).
\bibitem{Ijiri} Y. Ijiri, J. A. Borchers, R. W. Erwin, S. H. Lee, P. J. van der Zaag, and R. M. Wolf, Phys. Rev. Lett. {\bf 80}, 608 (1998).
\bibitem{Nogues2} J. Nogu\'{e}s, T. J. Moran, D. Lederman, I. K. Schuller, and K. V. Rao, Phys. Rev. B {\bf 59}, 6984 (1999).
\bibitem{Moran} T. J. Moran, J. Nogu\'{e}s, D. Lederman, and I. K. Schuller, Appl. Phys. Lett. {\bf 72}, 617 (1998).
\bibitem{Fitzsimmons} M. R. Fitzsimmons, C. Leighton, J. Nogu\'{e}s, A. Hoffmann, K. Kiu, C. F. Majkrzak, J. A. Dura, J. R. Groves, R. W. Springer, P. N. Arendt, V. Leiner, H. Lauter, and I. K. Schuller, Phys. Rev. B {\bf 65}, 134436 (2002).
\bibitem{Sano} M. Sano, S. Araki, M. Ohta, K. Noguchi, H. Morita, and M. Matsuzaki, IEEE Trans. Magn. {\bf 34}, 372 (1998).
\bibitem{Bae} S. Bae, J. H. Judy, W. F. Egelhoff, Jr., and P. J. Chen, J. Appl. Phys. {\bf 87}, 6650 (2000).
\bibitem{Leighton} C. Leighton, A. Hoffmann, M. R. Fitzsimmons, J. Nogu\'{e}s, and Ivan K. Shuller, Phil. Mag. B, {\bf 81}, 1927 (2001).
\bibitem{Fujii} T. Fujii, M. Takano, R. Katano, Y. Isozumi, and Y. Bando, J. Magn. Magn. Mater. {\bf 135}, 231 (1994).
\bibitem{Gota} S. Gota, M. Gautier-Soyer, and M. Sacchi, Phys. Rev. B {\bf 64}, 224407 (2001).
\bibitem{Fuller} M. Fuller, $\it{Methods of Experimental physics}$, vol. 24A, edited by C. G. Sammis, and T. L. Henyey, pp. 303-471, Academic press, Orlando (1987).
\bibitem{Vasquez} M. Vasquez-Mansilla, R. D. Zysler, C. Arciprete, M. Dimitrijewits, D. Rodriguez-Sierra, and C. Saragovi, J. Magn. Magn. Mater. {\bf 226-230}, 1907 (2001).
\bibitem{Hu} J. Hu, G. Jin, and Y. Ma, J. Appl. Phys. {\bf 94}, 2529 (2003).
\bibitem{Nowak} U. Nowak, K. D. Usadel, J. Keller, P. Miltenyi, B. Beschoten, and G. Guntherodt, Phys. Rev. B {\bf 66}, 014430 (2002).
\bibitem{Shang} C. H. Shang, G. P. Berera, and J. S. Moodera, App. Phys. Lett. {\bf 72}, 605 (1998).

\end{references}
\end{document}